\documentclass[twocolumn]{aastex62}

\newcommand{\quasar}{P352--15}

\shorttitle{Resolving The Powerful Radio-Loud Quasar at $z\sim 6$}
\shortauthors{Momjian et al.}

\begin{document}

\title{RESOLVING THE POWERFUL RADIO-LOUD QUASAR AT $\lowercase{z}\sim 6$}

\correspondingauthor{Emmanuel Momjian}
\email{emomjian@nrao.edu}

\author{Emmanuel Momjian}
\affiliation{National Radio Astronomy Observatory, P.O. Box O, Socorro, NM 87801, USA}

\author{Christopher L. Carilli}
\affiliation{National Radio Astronomy Observatory, P.O. Box O, Socorro, NM 87801, USA}
\affiliation{Astrophysics Group, Cavendish Laboratory, JJ Thomson Avenue, Cambridge CB3 0HE, UK}

\author{Eduardo Ba\~nados}
\altaffiliation{Carnegie-Princeton Fellow}
\affiliation{The Observatories of the Carnegie Institution for Science, 813 Santa Barbara St., Pasadena, CA 91101, USA}

\author{Fabian Walter}
\affiliation{{Max Planck Institut f\"ur Astronomie, K\"onigstuhl 17, D-69117, Heidelberg, Germany}}
\affiliation{National Radio Astronomy Observatory, P.O. Box O, Socorro, NM 87801, USA}

\author{Bram P. Venemans}
\affiliation{{Max Planck Institut f\"ur Astronomie, K\"onigstuhl 17, D-69117, Heidelberg, Germany}}

\begin{abstract}
We present high angular resolution imaging ($23.9 \times 11.3$\,mas, $138.6 \times 65.5$\,pc)
of the radio-loud quasar PSO~J352.4034$-$15.3373 at $z=5.84$ with the Very Long
Baseline Array (VLBA) at 1.54\,GHz. This quasar has the highest radio-to-optical flux density ratio at such a
redshift, making it the radio-loudest source known to date at $z \sim 6$. The VLBA observations presented here
resolve this quasar into multiple components with an overall linear extent of 1.62\,kpc ($0\rlap{.}{''}28$) and
with a total flux density of $6.57 \pm 0.38$\,mJy,
which is about half of the emission measured at a much lower angular resolution.
The morphology of the source is comparable with either a radio core with a one-sided jet, or a compact or
a medium-size Symmetric Object (CSO/MSO).
If the source is a CSO/MSO, and assuming an advance speed of $0.2c$, then the estimated kinematic age is
$\sim 10^4$\,yr.

\end{abstract}


\section{Introduction} 

The source PSO J352.4034$-$15.3373 (hereafter \quasar) has been identified as a luminous
quasar at $z = 5.84\pm 0.02$ \citep{ban18}. The optical spectrum shows the typical
broad emission lines of a high-redshift quasar, plus a possible associated
broad absorption line (BAL) system. Karl G. Jansky Very Large Array (VLA)
observations at 3\,GHz show that the quasar is associated with a radio source that has a 
flux density of  $8.20 \pm 0.25$\,mJy and a size $< 0.5''$.
At 1.4\,GHz, the flux density of the source is $14.9 \pm 0.7$\,mJy \citep{CON98}.
The source is also detected at lower frequencies in both the
TIFR GMRT Sky Survey (TGSS; \citealt{int17})
and the GaLactic and Extragalactic All-sky MWA Survey (GLEAM; \citealt{hw17}), with a flux density of around
100\,mJy at 200\,MHz. 
The implied rest frame 1.4\,GHz luminosity density is
$L_{\rm \nu,1.4\,GHz} = (4.5 \pm 0.2) \times 10^{27}$~W~Hz$^{-1}$ \citep{ban18}.

This quasar is the brightest of any known radio source at $z > 5.5$. 
With a radio loudness parameter of $R = f_{v,5\rm GHz}/f_{v,4400\rm A}$ $\gtrsim 10^3$, \quasar\ is
about an order of magnitude more radio loud than any other quasar at these redshifts
\citep{ban15,ban18}. The discovery of the quasar  is reported in the companion paper
\citep{ban18}
which presents its optical spectrum, its identification as a bright radio source using new VLA data as well as
available public surveys,
and its radio spectral energy distribution.

In this paper we present 1.54\,GHz Very Long Baseline Interferometry (VLBI)
imaging of this source.
These observations reveal a morphology that is consistent with a radio galaxy but with a linear
projected size of only 1.62\,kpc, or a radio core with a one-sided jet.
We discuss the Very Long Baseline Array (VLBA) result in the context of
quasar-mode feedback during the earliest formation of active galactic nuclei (AGN) and the most massive
galaxies. We adopt a flat cosmology with $H_0 = 70
\,\mbox{km\,s}^{-1}$\,Mpc$^{-1}$, $\Omega_M = 0.3$, and
$\Omega_\Lambda = 0.7$.  At the redshift of this quasar, 1\arcsec\
corresponds to 5.8\,kpc.

\section{Observations and Data Reduction} \label{obs}

The VLBI observations of the \quasar\ were carried out at 1.54\,GHz on
2018 January 23, using the VLBA of the Long Baseline Observatory (LBO).\footnote{The Long Baseline Observatory
is a facility of the
National Science Foundation operated under cooperative agreement by
Associated Universities, Inc.} Eight 32\,MHz data channel pairs were
recorded at each station using the ROACH Digital Backend and the
polyphase filterbank (PFB) digital signal-processing algorithm, both
with right- and left-hand circular polarizations, and sampled at two
bits. The total bandwidth was 256\,MHz centered at 1.54\,GHz. The
total observing time was 2~hr.

The VLBA observations employed nodding-style phase referencing using
the nearby calibrator J2327$-$1447 with a cycle time of 3.5\,min:
2.75\,min on the target and 0.75\,min on the calibrator. This calibrator
is at an angular distance of $0\rlap{.}{^\circ}7$ from the target
source. The accuracy of the phase calibrator position is important in
phase-referencing observations \citep {WAL99}, because it determines
the accuracy of the absolute position of the target and its
components, if any. The uncertainty in the position of the phase calibrator is
0.09\,mas in right ascension and 0.13\,mas in declination \citep{fey15}.
Phase referencing, as utilized in these
observations, is known to preserve absolute astrometric positions to
better than $\pm 0\rlap{.}^{''}01$ \citep{FOM99}. The calibrator
source 3C\,454.3 was used as a fringe finder and bandpass
calibrator. Amplitude calibration was performed using measurements of
the antenna gain and the system temperature of each station. The data
were correlated with the VLBA DiFX correlator \citep{DEL11} in
Socorro, New Mexico, with 2~s correlator integration time. 

Data reduction and analysis were performed using the Astronomical
Image Processing System (AIPS: \citealt{G2003}) following standard VLBI
data reduction procedures. The phase reference source was
self-calibrated, and the solutions were applied on the target. The
continuum emission from the target was then deconvolved and imaged
using a grid weighting between natural and uniform (Robust=0 in AIPS
task IMAGR).

\section{Results and Analysis} \label{resultsandanalysis}

Figure~1 shows the 1.54\,GHz image obtained on the target source
\quasar\ with the VLBA at an angular resolution of $23.9 \times
11.3$~mas ($138.6 \times 65.5$\,pc at $z=5.84$) with a position angle of
P.A.=$8^{\circ}$. The rms noise in the image is 67~$\mu$Jy~beam$^{-1}$.
The observing frequency of 1.54\,GHz corresponds
to a rest frame frequency of 10.53\,GHz. The continuum emission from
\quasar\ is resolved into three main structures with an overall
linear extent of 1.62\,kpc ($0\rlap{.}{''}28$) at P.A.=
$119^{\circ}$. From east to west, we denote these structures as E
(east), C (center), and W (west). The E structure is dominated by a
single component, while the C and W structures are resolved into
multiple components. We performed Gaussian fitting on these structures
with a single component on E, and two components on each of C and W. Table\,1 lists the Gaussian fit parameters of the
continuum features in \quasar. Column (1) denotes the names of the
components. Column (2) lists the relative distance of each component
with respect to the stronger of the two center sources, C1, and column (3) lists their flux
densities. Column (4) gives the nominal or maximum (upper limit)
deconvolved sizes for the major and minor axes of each component, and column (5) reports the
corresponding intrinsic brightness temperature values (i.e., at the rest frame frequency of 10.53\,GHz), or their lower
limits. The components listed in Table~1 are marked in Figure\,1.
The absolute positional accuracy of the phased-referenced VLBA image (Figure\,1) and the VLBA detected sources listed in Table\,1
is expected to be better than $\pm 0\rlap{.}^{''}01$ (see Section \ref{obs} for further details).
We note that Gaussian component fitting, as listed in
Table\,1, provides a convenient measure of the source structure even if
they do not necessarily represent discrete physical structures.

The optical position of  \quasar\ is taken from the Pan-STARRS1 (PS1) catalog \citep{cham16}. Positions in the
PS1 catalog are calibrated using Gaia Data Release 1 (see \citealt{mag16}). We verified the astrometry of the PS1 catalog
by comparing the coordinates of 70 objects within 5\arcmin\ of \quasar\ with positions listed in Gaia Data Release~2 (Gaia
DR2, \citealt{gaia16,gaia18}). We find a mean shift in right ascension and declination of $3\pm5$\,mas and $5\pm6$\,mas,
respectively. We conclude that the differences in the position of objects near the quasar in the PS1 and Gaia DR2 catalogs
are negligible, and we take the PS1 coordinates of the quasar for its optical position. The optical quasar coordinates are
R.A.\,(J2000)=\,23$^{\rm h}$29$^{\rm m}$36.8363$^{\rm s}$, Decl.\,(J2000)=\,$-$15$^\circ$20$^\prime$14\farcs460, with
an uncertainty of $\pm64$\,mas in both right ascension and declination.

The total flux density of \quasar measured with the VLBA at 1.54\,GHz is $6.57 \pm
0.38$\,mJy. The flux density measured with the VLA at 1.4\,GHz is
$14.9 \pm 0.7$\,mJy \citep{CON98}, and at 3\,GHz is $8.20 \pm
0.25$\,mJy \citep{ban18}, resulting in a spectral index
between 1.4 and 3\,GHz of $\alpha^{3.0}_{1.4}=-0.78$ (adopting $S
\sim \nu^{\alpha}$). This suggests that the total flux density of the
source at 1.54\,GHz would be 13.83\,mJy. Therefore, with the resolution
of the VLBA, we are only recovering about half of the total flux density of
the source.

These spatially resolving observations can be used to calculate at
least a minimum pressure in the radio source, through the standard
synchrotron minimum energy arguments. We use the equations in \citet{m80},
and assume a proton-to-electron number ratio and a filling
factor for the synchrotron emitting fluid of unity and a minimum and
maximum frequency for the emission of 0.01\,GHz to 10\,GHz. The typical
observed surface brightness for the resolved components in Table\,1 is
$\sim 4$\,Jy\,arcsec$^{-2}$, and the line-of-sight depth, assuming
rotation symmetry, is about 0.1\,kpc. The implied minimum pressure is
$\sim 4\times 10^{-7}$\,dyne\,cm$^{-2}$, and the corresponding magnetic field
strength is $ \sim 3.5$\,mG. For comparison, the hot spots of large ($\sim 100$\,kpc)
classical double radio galaxies, such as Cygnus A, have fields about
an order of magnitude smaller and pressures two orders of magnitude
smaller \citep{cb96}.

\section{Discussion}

We have detected the $z=5.84$ quasar \quasar\ at 1.54~GHz with the
VLBA at mas resolution. The observations show that the radio source is
comprised of several components with a maximum angular separation of
$0\rlap{.}{''}28$ (1.62\,kpc). The total flux density measured with the VLBA
is $6.57 \pm 0.38$\,mJy, suggesting that about half of the radio emission
associated with the quasar is resolved out, or below our detection
threshold. \quasar, which is believed to be near the end of the epoch of reionization,
would be an excellent candidate for H{\footnotesize I} 21\,cm absorption experiments to detect the
neutral intergalactic medium (IGM) at such high redshifts \citep{cgo02,FL02,GM17}. Knowledge of the source
structure, as presented herein, is critical for both identifying
potential candidates for neutral hydrogen absorption searches and for
subsequent interpretation of the results.

We consider two possible interpretations of the morphology seen in the VLBA image of \quasar\ (Figure~1):
(i) a radio core with a one-sided jet, and (ii) a classic, but compact, Fanaroff--Riley Class II (FRII) source.
With the available data, both scenarios remain viable, even though the proximity of the optical position to the
E component makes the core with a one-sided jet interpretation more likely.
Multifrequency VLBI observations are needed to measure the
spectra of the various components seen in \quasar\ in order to identify a flat spectrum core in
this source and determine its exact nature.

\subsection{A Radio Core with a One Sided Jet}
In the scenario that the source is a radio core with a one-sided jet,
component E (see Figure~1 and Table 1) is the core, and the C and W components are part of a one-side jet
structure. The E component is within $1\sigma$ of the optical position of the quasar. 
If the source is indeed a one-sided jet, \quasar\ becomes the most distant source in
which to study relativistic jet expansion.

The maximum jet apparent proper motion of $\sim 0.2$\,mas\,yr$^{-1}$ is predicted
at the highest redshifts based on a maximum jet Lorentz factor of 25 \citep{kel04,FR15}.
VLBI studies of the radio blazar J1026+2542 at $z  = 5.27$
\citep{FR15} and the quasar J2134$-$0419 at $z=4.3$ \citep{PE18} are at
least consistent with this prediction. The maximum jet apparent proper motion limit noted above
corresponds to 1\,mas in a 5 year time baseline. Such a proper motion could be measured in \quasar\ via
sensitive astrometric VLBI observations.
Furthermore, if a proper motion of substantially greater than $0.2$\,mas\,yr$^{-1}$ is observed in \quasar,
then,
according to the analysis presented by \citet{kel04}, this would be the jet with the highest
Lorentz factor. Overall, currently there are only a few radio sources with jets known at $z> 3.5$ (e.g., \citealt{FR15}), hence it is crucial to have more such sources in order to study and test
the apparent proper motion -- redshift relation at these high redshifts.

\subsection{A Compact FRII Radio Morphology: A CSO/MSO}
In the scenario that the source is a compact FRII radio galaxy, the radio core will be in the C region, while the E
and the W components will be the edge-brightened radio lobes separated by 1.62\,kpc.
The C region is within $3\sigma$ of the optical position of the quasar.
Contrary to typical FRII sources that span linear extents of 10s to 100s of kpc, this source has an extent of only 1.62\,kpc, 
placing this high-$z$ radio-loud quasar near the transition between the category of compact symmetric objects: CSO
= FRII radio sources with sizes $0.01 - 1$\,kpc, and medium-size symmetric objects: MSO = FRII radio sources with sizes $1 - 10$\,kpc \citep{RTXPWP96,OC98,CO02}.

CSOs and MSOs are considered to be young radio galaxies with powerful
radio jets that are strongly confined by the ISM in the center of the
host galaxy \citep{CO02}. \citet{TMPR00} present the most
detailed VLBI imaging analysis of a small sample of CSOs with sizes
ranging from 20\,pc to 100\,pc. They measure projected advance speeds for
the hot spots around $0.2c$, and source ages of a few hundred to 1000
years. The minimum pressures in the radio hot spots are of the order of
$10^{-4}-10^{-6}$\,dyne\,cm$^{-2}$ \citep{RTPW96}.

The most distant CSO found to date is the $z=6.12$ quasar, J1427+3312
\citep{FR08,MOM08}. The difference is that \quasar\ is about 10
times more luminous in the radio and about 10 times bigger in
projected linear size, than J1427+3312. At a lower redshift of $z=5.19$, the
bright radio galaxy TN~J0924-2201 is an order of magnitude more luminous in the radio than \quasar. However, TN~J0924-2201
is larger in extent by a factor of $\sim 5$ compared to \quasar\ \citep{vb99}. This may suggest that \quasar\ is
embedded in a denser environment and/or a younger radio source when compared to TN~J0924-2201.

Using the minimum pressure calculated in Section \ref{resultsandanalysis},
we can compute a hot spot advance speed
assuming ram-pressure confinement. In this case, $\rho v^2 \sim
P_{min}$, where $\rho$ is the mass density of the external
medium. Unfortunately, we do not have an estimate of $\rho$.  If we were to
adopt the average density of the ISM in the plane of the Milky Way of
$n \sim 1$ cm$^{-3}$, then the implied advance speed is $\sim 0.02c$, which is comparable
to what may be typical for CSOs \citep{RTXPWP96}.
In some coeval starburst-QSO systems at $z \sim 6$, an analysis of CO excitation levels
suggests much higher densities, approaching $10^5$\,cm$^{-3}$ (see
\citealt{cw13}). The ISM density on sub-kpc scales in nearby
AGN can be of the order of 100\,cm$^{-3}$, or greater \citep{mor13}.
Higher densities imply a lower advance speed and larger age. However,
if the jets are emerging along the axis of a disk galaxy,
relevant densities may correspond to the galaxy halo, and hence could
be much lower, implying a higher advance speed and a younger age.

Assuming an advance speed of $\sim 0.02c$, as derived using the minimum pressure argument, and the average density
of the ISM in the plane of the Milky Way, then the age of the source is $\sim 10^5$\,years. However, if the advance speed
in \quasar\ is similar to that measured in low redshift CSOs through multi-epoch VLBI imaging
\citep{TMPR00}, which is $\sim 0.2c$, then the derived kinematic age of the source is $\sim 10^4$\,years.

We emphasize here that the implied source advance speed and age cannot be robustly determined until either 
the ISM density is determined or until a direct measurement of hot spot motions is made.
The former might be possible with ALMA observations
of millimeter line or dust emission. The latter would likely require 
multi-epoch VLBI observations at very high resolution and sensitivity.
For instance, if the advance speed is $v_a=0.2c$ on the plane of the sky, then the separation between the hot
spots increases by $0.4c$, and in one year the hot spots will separate by $\sim 0.12$\,pc. The angular size distance at the
redshift of the quasar is 1195.3\,Mpc; therefore, the separation between the
hot spots would increase at $\sim 20 \times (v_a/0.2c)$\,$\mu$arcsec\,yr$^{-1}$.
\citet{TMPR00} measured angular separation rates ranging between
$\sim$ 5 and 80\,$\mu$arcsec\,yr$^{-1}$ in low-redshift CSOs with the VLBA at 15\,GHz through multi-epoch
observations spanning up to 5 years. 
However, these low-redshift CSOs and their components are significantly brighter compared to \quasar\ and its
components. Therefore, if the source is a CSO/MSO, the apparent proper motion will be much less compared to
the one-sided jet scenario. Such measurements will likely require
a future VLBI array that would include ultra-sensitive elements, such as
the Next Generation Very Large Array (ngVLA\footnote{\url{ngvla.nrao.edu}}) and the Square Kilometre Array
(SKA\footnote{\url{www.skatelescope.org}}).

The extreme redshift and strong radio emission from \quasar\ presents a
unique potential to constrain feedback processes in the first
galaxies. Radio jets are hypothesized to play a defining role in
massive galaxy formation, through negative or ``radio mode feedback'',
in which the radio jets evacuate the ISM of the host galaxy \citep{sil13}.
Because CSOs/MSOs are highly confined sources, they are
thought to have higher conversion efficiency of jet kinetic energy to
radio luminosity. This process has been studied in detail on
pc-scales in lower-redshift CSOs, such as 4C12.50 \citep{mor13,mor15}. 
A recent possible analogy on larger scales is 3C\,298
at $z = 1.4$ \citep{vay17}. In this case, the radio source is
10 times larger than \quasar. Still, imaging of the ionized and
molecular gas and dust in the host galaxy implies that a strong outflow is
being driven by either the expanding radio source and/or a wind from
the luminous AGN.

We can only speculate at the present that \quasar\ will exhibit 
clear signs of radio-mode feedback, although an interesting
similarity between this quasar and 3C\,298 is a likely associated BAL
system in the optical spectrum of \quasar\ \citep{ban18}. BALs are thought to be
indicative of strong nuclear outflows. Further ALMA and VLBI observations
are planned to determine the ISM properties of the host galaxy
and the radio source properties in great detail. 

\acknowledgments
The authors are grateful to the anonymous referee for their valuable comments that helped improve the 
content of this work.
The National Radio Astronomy Observatory is a facility of the National Science Foundation operated
under cooperative agreement by Associated Universities, Inc. F.W.\ and B.P.V.\ acknowledge
funding through ERC grants ``Cosmic Dawn'' and ``Cosmic Gas.'' This work has made use of data from
the European Space Agency (ESA) mission
{\it Gaia} (\url{https://www.cosmos.esa.int/gaia}), processed by the {\it Gaia}
Data Processing and Analysis Consortium (DPAC,
\url{https://www.cosmos.esa.int/web/gaia/dpac/consortium}). Funding for the DPAC
has been provided by national institutions, in particular the institutions
participating in the {\it Gaia} Multilateral Agreement.

\facilities{VLBA.}

\clearpage
\begin{figure}
\epsscale{1}
\plotone{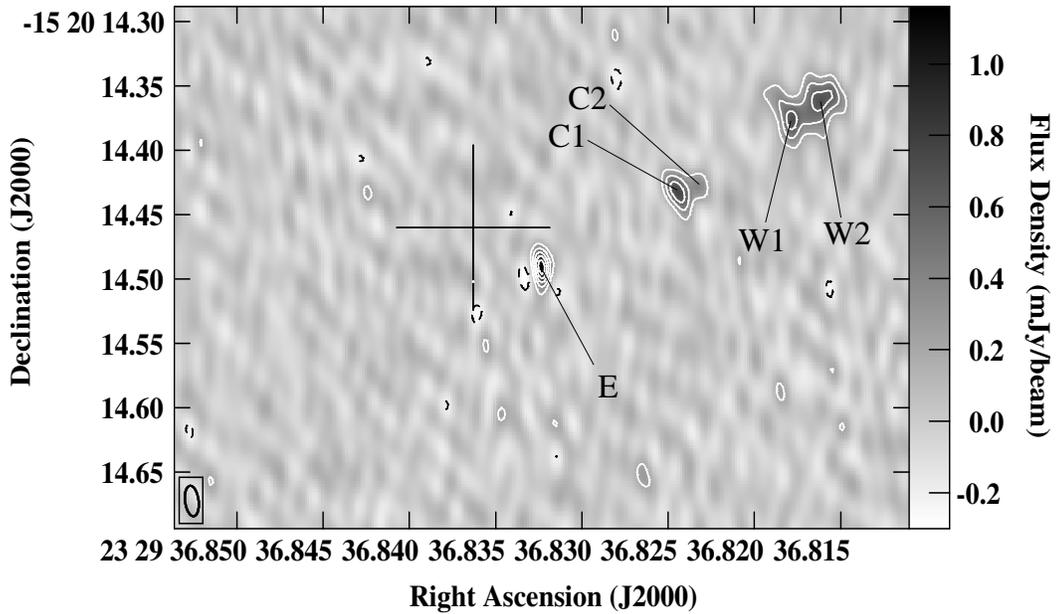}
\caption
{VLBA continuum image of the $z=5.84$ QSO \quasar\ at 1.54\,GHz and $23.9 \times
11.3$~mas resolution (P.A.=$8^{\circ}$). The contour levels are at $-3$, 3, 6, 9, 12, and 15 
times the rms noise level, which is
67~$\mu$Jy~beam$^{-1}$. The gray-scale range is indicated by the step wedge at the right side of the image.
The plus sign denotes the optical position of the quasar: R.A.\,(J2000)=\,23$^{\rm h}$29$^{\rm
m}$36.8363$^{\rm s}$, 
Decl.\,(J2000)=\,$-$15$^\circ$20$^\prime$14\farcs460 with an uncertainty of $\pm 64$\,mas
in both right ascension and declination.
\label{fig:vlba}}
\end{figure}

\begin{deluxetable}{cccccrr}
\tablecolumns{6}
\tablewidth{0pc}
\tablecaption{Gaussian Fits to the Continuum Components of \quasar}
\tablehead{
\colhead{} &  & \colhead{Relative Position\tablenotemark{a}}&
\colhead{Flux Density} &
\colhead{Deconvolved Size} & \colhead{T$_b \times 10^7$\tablenotemark{b}} \\
\colhead{Source} & \colhead{} & \colhead{(mas)} & 
\colhead{(mJy)} &\colhead{(mas)}& \colhead{(K)} \\
\colhead{(1)} & & \colhead{(2)}& \colhead{(3)} & \colhead{(4)}
& \colhead{(5)}}
\startdata
E  & &  114E, 57S  & $1.17 \pm 0.07$ & $ 5.3~\times < 6.0 $&  $ > 13.02 \pm 0.78 $  \\
C1 & &   0,   0    & $1.08 \pm 0.14$ & $18.0~\times < 6.9 $&  $ > 3.07   \pm 0.39 $ \\
C2 & &  16W, 6N  & $0.41 \pm 0.07$ & $14.5~\times < 15.8 $&  $ > 0.63   \pm 0.11 $ \\
W1 & &  92W, 59N & $1.37 \pm 0.21$ & $34.2~\times 10.3   $&  $   1.37   \pm 0.21 $ \\
W2 & &  119W, 70N & $2.54 \pm 0.26$ & $30.7~\times 16.9   $&  $   1.73   \pm 0.18 $ \\
\enddata
\tablenotetext{a}{The position (0,0) is $\alpha\rm{(J2000)}=23^{\rm
h} 29^{\rm m} 36\rlap{.}^{\rm s} 8245$, $\delta\rm{(J2000)}=-15^{\circ}20{'}14\rlap{.}{''}433$.}
\tablenotetext{b}{At the rest frame frequency of 10.53\,GHz.}
\end{deluxetable}

\end{document}